\documentstyle[12pt]{article}
\begin{document}

\begin{titlepage}
\hfill CERN-TH.7029/93 \\
\vspace{1mm}
\hfill MS-TPI-93-08 \\

\begin{centering}
\vfill
{\LARGE \bf Comparison of Monte Carlo Results \\
            for the 3D Ising Interface Tension \\
            and Interface Energy \\
            with (Extrapolated) Series Expansions \\
            }

\vspace{2cm}
{\bf M.\ Hasenbusch$^1$ and K.\ Pinn$^2$ } \\[6mm]

\vspace{1cm}
{\em $^1
\, $Theory Division, CERN,\\ CH-1211 Geneva 23, Switzerland}

\vspace{0.3cm}
{\em $^2$
\, Institut f\"ur Theoretische Physik I, Universit\"at M\"unster, \\
   Wilhelm-Klemm-Str.\ 9, D-48149 M\"unster, Germany \\}

\vspace{2cm}
{\bf Abstract} \\
\end{centering}
\vspace{0.2cm}

We compare Monte Carlo results
for the interface tension and interface
energy of the 3-dimensional Ising model
with Pad\'e and inhomogeneous differential
approximants of the low temperature series
that was recently extended by Arisue to $17^{\rm th}$ order in
$u=\exp(-4\beta)$. The series is expected to suffer from
the roughening singularity at $u \approx 0.196$.
The comparison with the Monte Carlo
data shows that the Pad\'e and inhomogeneous differential
approximants fail to improve the truncated series
result of the interface tension and the interface energy
in the region around the roughening transition.
The Monte Carlo data show that the specific heat displays
a peak in the smooth phase. Neither the truncated series nor the
Pad\'e approximants find this peak.
We also compare Monte Carlo data for the energy of the
ASOS model with the corresponding low temperature series
that we extended to order $u^{12}$.

\vspace{0.3cm}\noindent

\vfill \vfill
\noindent
CERN-TH.7029/93\\
MS-TPI-93-08\\
October 1993
\end{titlepage}
%
\newcommand{\nc}{\newcommand}
\nc{\be}{\begin{equation}}
\nc{\ee}{\end{equation}}
\nc{\bea}{\begin{eqnarray}}
\nc{\eea}{\end{eqnarray}}
\nc{\rbo}{\raisebox}
\nc{\cH}{{\cal H}}
\nc{\RR} {\rangle \! \rangle}
\nc{\LL} {\langle \! \langle}
\nc{\rmi}[1]{{\mbox{\small #1}}}
\nc{\eq}{eq.~}
\nc{\nr}[1]{(\ref{#1})}
\nc{\ul}{\underline}
\nc{\cM}{{\cal M}}
\nc{\mc}{\multicolumn}
\thispagestyle{empty}
\newpage

\section{Introduction}
\label{SECintro}

 Properties of the interface separating coexisting phases
 of the 3-dimensional Ising model have found continuous
 interest in the literature.

 The Ising interface undergoes
 a roughening transition at an inverse temperature
 $\beta_r = 0.4074(3)$ \cite{martinthesis}
 that is nearly twice as large as the bulk
 transition coupling $\beta_c= 0.221652(3)$ \cite{betac}.

 The roughening transition is believed to be of the
 Kosterlitz-Thouless (KT) nature \cite{KT}, with the
 surface free energy having an essential singularity
 at $\beta_r$ of the type
\be\label{fsing}
 f_{\rm sing} \sim \exp[ - A (\beta - \beta_r)^{-1/2} ] \, .
\ee
 Though the free energy and all its derivatives with respect
 to $\beta$ stay finite at the roughening point, one has
 to expect that low temperature series for interface properties
 suffer from the transition.

 The first low temperature expansion of the 3D Ising interface
 tension $\sigma$ was given by Weeks et al.\ to $9^{\rm th}$
 order in the variable $u=\exp(-4\beta)$.
 Shaw and Fisher \cite{ShawFisher} analyzed
 the series with the help of Pad\'e and
 inhomogeneous differential approximants.
 They claimed that the Pad\'e and differential
 approximants allow to compute
 the surface tension accurately for temperatures below the
 roughening point.

 Recently, Arisue put forward the series to $17^{\rm th}$ order
 in $u$ \cite{ArisueNewSeries}.
 It is interesting to note that the coefficients of the series change
 their sign at order 13. This behavior does not come unexpected and
 confirms that the roughening transition of the Ising interface is of
 Kosterlitz-Thouless type: Expanding eq.~(\ref{fsing}) in the variable $u$,
 one also obtains a series with coefficients that change
 their sign at a certain order. The order where the change
 of sign happens depends on the non-universal parameters
 $A$ and $\beta_r$.

 A completely independent method to study
 the Ising interface is the Monte Carlo method
 \cite{Binder}. For recent Monte Carlo work
 on the 3D Ising interface see, e.g.,
 \cite{Klessinger,BergHans,ours,Sanie,direct,ItoPhysA}
 and references cited therein.

 In \cite{ours}, we reported on a numerical study of
 properties of the Ising interface over the whole
 range from low temperatures up to the bulk
 critical point. In particular we determined the surface energy
 and, by integration over $\beta$, also the
 surface free energy and surface tension.

 In this paper, we give a more detailed account
 of the numerical results and compare them
 with Pad\'e and inhomogeneous differential
 approximants for the extended  low temperature series.
 In order to demonstrate that the disagreements of
 series and numerical data are not due
 to finite size effects, we provide data
 for various lattice extensions and demonstrate
 that the systematic errors in the determination
 of surface energy and surface tension are under
 control.

 The ASOS (absolute value solid-on-solid)
 model in two dimensions is an approximation of a
 phase separating surface in a 3-dimensional Ising model.
 The approximation neglects overhangs of the surface and
 fluctuations of the bulk phases. For temperatures being
 low compared to the bulk critical temperature of the Ising
 model this approximation should describe
 the Ising surface quantitatively rather well.
 We extended the existing low temperature series
 for the ASOS model to $12^{\rm th}$ order and
 compared its Pad\'e approximants with Monte
 Carlo results that we obtained with the
 VMR cluster algorithm \cite{VMRthegang}.

 This paper is organized as follows. In section \ref{SECmodels}
 we introduce the notation for the 3D Ising model
 and its ASOS approximation.
 We also define the interface tension
 and discuss the problem of how to obtain it
 from simulations of finite lattices.
 The algorithms used for the
 Monte Carlo simulations are specified in
 section \ref{SECalgorithms}. In
 section \ref{SECcomparison} we quote the results and
 compare them with extrapolations of low temperature
 series. Conclusions are given in section \ref{SECconclusions}.

\section{The Models and Quantities Studied}
\label{SECmodels}

\subsection*{3-Dimensional Ising Model}

 We consider a simple cubic lattice with extension
 $L$ in $x$- and $y$-direction and with extension $t$
 in $z$-direction. The Ising model is defined by the partition function
\be
 Z = \sum_{ \{ \sigma_i = \pm 1\} } \exp(-\beta H) \, .
\ee
 The Hamiltonian $H$ is a sum over nearest neighbor contributions,
\be
 H= - \sum_{<i,j>} k_{ij} \sigma_i \sigma_j  \, .
\ee
 The interaction energy is normalized such
 that $\beta = J / k_{\rm B}{\tilde T}$,
 where $k_{\rm B}$ denotes Boltzmann's constant, $J$ is the interaction
 energy, and $\tilde T$ is the temperature.

 The lattice becomes a torus by defining that
 the uppermost plane is regarded as the lower neighbor plane
 of the lower-most plane. An analogous identification is done for
 the other two lattice directions. For the Ising spins
 $\sigma$ we use two different boundary conditions:
 {\em Periodic} boundary conditions are defined by letting
 $k_{ij}=1$ for all links $< \! i,j \!>$ in the lattice.
 To define {\em antiperiodic} boundary conditions in $z$-direction, we
 also set $k_{ij}=1$ with the exception of the
 links connecting the uppermost plane
 with the lower-most plane. For these links we set $k_{ij}=-1$.

 Let us  define
\be
 G = - \left( \ln Z_{a} - \ln Z_{p} \right) \, ,
\ee
 where the subscript $a$ ($p$) means antiperiodic (periodic)
 boundary conditions.
 The {\em surface tension} can be defined as the limit
\be
 \sigma =
  \lim_{t \rightarrow \infty} \lim_{L \rightarrow \infty} \frac{G}{L^2} \, .
\ee
 With numerical simulations only a rather limited
 range of $L$ and $t$ values is accessible. Hence a careful discussion
 of finite size effects is needed. Let us express the partition functions
 of the periodic and antiperiodic Ising system in terms of the
 transfer matrix $T$. The antiperiodic  boundary
 conditions are represented by
 a spin-flip operator $P$ that flips the sign
 of all spins in a given $z$-slice.

 The partition function of the periodic system is given by
 $ Z_{p} = {\rm Tr} T^t $, while the partition function
 of the antiperiodic system is given by
 $ Z_{a} = {\rm Tr} T^t P$.
 The operators $T$ and $P$ commute and thus
 have a common set of eigenfunctions.
 Say the eigenvalues of $T$ are $\lambda_i$ and
 those of $P$ are $p_i$. The possible values of $p_i$ are $1$ and $-1$.
 The partition functions take the form
 $ Z_{p} =  \sum_i \lambda_i^t $
 and  $ Z_{a} =  \sum_i \lambda_i^t p_i $.

 Let us consider the ratio of the partition functions
 in the broken phase of the model. Here
 the two largest eigenvalues $\lambda_{0s} $ and $\lambda_{0a}$
 are much larger than the other eigenvalues \cite{PrivmanFisher}.
 (The subscripts $s$ and $a$ label
 eigenfunctions with $p=1$ and $p=-1$, respectively.)
 Hence the ratio of the two partition functions can be well approximated by
\be
 \frac{Z_{a}}{Z_{p}} \simeq \frac { \lambda_{0s}^t - \lambda_{0a}^t}
                             { \lambda_{0s}^t + \lambda_{0a}^t} \, .
\ee
 The corrections are of order $(\lambda_{1s} / \lambda_{0s})^t $.
 This means that the extension $t$ of the lattice in $z$-direction has to
 be much larger than the bulk correlation length $\xi$.
 For $\xi_{0a} \!>\!> t$
 ($\xi_{0a} = -1/\ln(\frac{\lambda_{0a}}{\lambda_{0s}})$
 is the tunneling correlation length)
 we can write
\be
 \frac{Z_{a}}{Z_{p}} \simeq
 \frac{t}{2} (1-\frac{\lambda_{0a}}{\lambda_{0s}}) \, .
\ee
 Notice that within this approximation
 the derivative of $G$ with respect to $\beta$
 does not depend on $t$.

 According to this discussion, if $ \xi \!<\!< t \!<\!< \xi_{0a} $
 then already for
 finite $L$ a surface free energy is rather well defined by
\be
 F_s \simeq G + \ln t \, .
\ee

 Phenomenologically one can interpret this situation as follows: \\
 The lattice is short enough that the creation of interfaces in the system
 with periodic boundary conditions is sufficiently suppressed while for
 the system with antiperiodic boundaries only the interface induced by the
 boundary conditions is present. On the other hand the extension of the
 lattice is large enough not to restrict the fluctuations of the interface.

 In order to discuss the $L$ dependence of the surface free energy
 a model for the surface is needed.

 Theoretical studies of the interface are based on
 the SOS (solid-on-solid) approximation
 which essentially neglects overhangs and bulk fluctuations.
 SOS models predict that the roughening transition is of
 the Kosterlitz-Thouless type \cite{KTSOS}.
 That the Ising interface at the roughening point is indeed
 in the same universality class as various SOS models was
 demonstrated by Hasenbusch using a renormalization group
 matching procedure \cite{martinthesis}. For large $\beta$
 there are also rigorous results from linked cluster expansions.
 Borgs and Imbrie have shown \cite{Borgsetal}
 for sufficiently large $\beta$, i.e.,
 when the interface is smooth, that
\be
 F_s \simeq  \sigma \, L^2 \, .
\ee
 It is believed that this result holds for all $\beta > \beta_r$.

 A model for the interface in the rough phase is
 the capillary wave model \cite{PrivmanCapillary}.
 In its quadratic approximation
 it essentially states that the infrared fluctuations of
 the interface are massless Gaussian. This assumption has
 been verified numerically in a number of cases, see e.g.\ \cite{ours}.
 The massless Gaussian dynamics
 leads to the following finite $L$ behavior of the surface free energy
 in the rough phase \cite{brezin,munster,bunk,cas}:
\be\label{roughfs}
 F_s \simeq C_s + \sigma \, L^2 \, .
\ee
 Gelfand and Fisher \cite{gelfand}
 predicted in addition a logarithmic dependence
 of the surface free energy on the lattice size. They  did not
 take into account a prefactor  $L$ in the partition function
 that arises when the average position of the interface
 is fixed via a $\delta$-function.

 At the roughening transition one has still a Gaussian fixed point,
 however, with logarithmic
 corrections. Hence eq.~(\ref{roughfs})
 should be still valid for sufficiently large $L$.

 It is difficult to compute free energies directly by
 Monte Carlo (however, cf.\ \cite{direct}).
 But the derivative of $G$ with respect to $\beta$ is
 a quantity that can be computed by Monte Carlo:
\be\label{def_surface_energy}
 \frac {\partial G}{\partial \beta} =
 \langle H \rangle_{a} - \langle H \rangle_{p}
 \equiv E_s \, .
\ee
 $G$ can then be obtained by integration over $\beta$:
\be
 G(\beta) = G(\beta_0) + \int_{\beta_0}^{\beta}
 d \beta' \,  E_s(\beta') \, ,
\ee
 where $\beta_0$ is arbitrary.
 In the case that there is only one interface in the system,
 $E_s$ is the {\em surface energy}. The surface energy per
 area is defined as
\be
 \epsilon_s = E_s/L^2 \, .
\ee

 In \cite{ours} we used the method of `integration over $\beta$'
 to determine the surface free energies for a wide range
 of temperatures, for $L=8,16,32,64$.
 We found that the surface free energy can be fitted accurately with
 eq.~(\ref{roughfs}).
    We thus identify the coefficient $\sigma$ in front of the
 factor $L^2$ with the {\em surface tension}.

\subsection*{ASOS Model}

 The ASOS model is the solid-on-solid approximation of an
 interface of a 3D simple cubic lattice Ising model
 (on a (001)-lattice plane). It lives on a {\em two}-dimensional
 square lattice of size $L$ by $L$.
 The partition function is
\be
 Z_{ASOS} = \sum_{h} \exp \left(- \beta \sum_{<ij>}
          |h_i-h_j| \right) \, ,
\ee
 with $h_i$ integer. We shall study the quantity
\be
 \epsilon_{ASOS} \equiv E_{ASOS}/L^2 = <\sum_{<i,j>} |h_i - h_j|> / L^2 \, .
\ee
 In the ASOS limit of the Ising model the inverse temperatures of
 the two models are related by
\be
 \beta_{\rm Ising} = \beta_{ASOS} / 2 \, .
\ee
 In the same limit, there is the following relation between
 the surface energies:
\be
 E_{s,\rm Ising} = 2 + 2 \, E_{ASOS}
\ee
 for corresponding $\beta$ values.
 The most recent estimate for the roughening coupling of the
 ASOS model is $\beta_{r,ASOS} = 0.8061(3)$ \cite{matching}.

\section{Monte Carlo Algorithms}
\label{SECalgorithms}

\subsection*{Ising Model}

\subsubsection*{Cluster Algorithm for the Ising Interface}

For the production of the Monte Carlo results of ref.\ \cite{ours}
and for part of the new results to be presented below
we employed the Ising interface cluster algorithm of Hasenbusch
and Meyer \cite{InterfaceAlgorithm}. A detailed description
of this algorithm can be found in \cite{ours}.
There we computed besides the
surface energies also nonlocal quantities like the
surface thickness and block spin correlation functions.
The cluster algorithm proved to be very efficient for
this purpose.

\subsubsection*{Local Demon Algorithm for the Ising Model}

When the focus is on the determination of the energy with
antiperiodic boundary condition, it is helpful to use
a local algorithm instead of the cluster algorithm.
It is much easier to adapt a local Monte Carlo algorithm
for vectorization, parallelization or multi-spin coding.

It turned out that the energy of the system with antiperiodic
boundary conditions does almost not couple to the slow modes
of the local algorithm.

For the update of the 3-dimensional Ising model we therefore
also used a micro-canonical demon algorithm
\cite{creutz83,creutz84,creutz86} in combination with
a particularly efficient canonical update \cite{kari93} of the demons.
The algorithm is implemented using the
multi-spin coding technique \cite{creutz84,creutz86}.
Every bit of a computer word carries one Ising spin. In order to avoid
restrictions of the geometry of the systems 32 independent
systems are run in parallel.
The demon carries three levels carrying the energies 4, 8 and 16.
The number of demons is chosen to be equal to the number of lattice sites.

The simulation is done by performing a cycle of 5 groups, where each
group consists of a {\em microcanonical} update of the spin demon
system and a {\em translation} of the demon layer with energy 4,8,
or 16 (alternating). Each group is finished by updating the
demons with energy 4. The hole cycle is completed by updating
the demons with energy 8.

\noindent
 The canonical update of a demon layer
 consists of the following steps \cite{kari93}:

\begin{enumerate}
 \item   calculate the total number $N_{\rm old}$
         of demons in that layer carrying energy
 \item   calculate a new total number $N_{\rm new}$
         of demons with energy  according to the
         probability  $p_{E} = n_E \exp(-\beta E) /Z_{D}$,
         with $Z_{D} = \sum_{E} n_E \exp(-\beta E)$,
         where $n_E$ is the number
         of demon states having the total energy $E$
 \item   {\em if} $N_{\rm new} > N_{\rm old}$ {\em then} pick randomly
         $N_{\rm new}-N_{\rm old}$
         empty demons and fill them, {\em else}
         pick randomly $N_{\rm old}-N_{\rm new}$ occupied
         demons and clear them
\end{enumerate}

 This is a valid update of the total demon system
 that provides new total energies with a heat bath distribution.
 Random numbers are only needed for the selection of the
 $|N_{\rm old}-N_{\rm new}|$ demons which change their energy
 and one for the selection of the new total energy.
 Notice that $|N_{\rm old}-N_{\rm new}|$ is only of the order of the
 square root of all the demons.

 Combined with shifts and translations of the demons such an update
 of the demons should be almost as good as a heat bath for every single
 demon.

 In table \ref{tabspeed} we quote the performance of our
 algorithm on various workstations. For comparison we also
 cite the performance data of algorithms of other authors
 on different supercomputers in table \ref{tabspeedcomparison}.

\subsection*{VMR Algorithm for the ASOS Model}

 The results for the energy of the ASOS model cited in this paper
 were obtained with the help of the VMR (valleys-to-mountains-reflection)
 algorithm introduced in \cite{VMRthegang}. The basic idea
 of this algorithm is to define valleys and mountains by cutting
 the configuration of height variables with a reflection plane.
 The flip of a cluster of spins can be regarded as reflecting
 a valley to a mountain (or vice versa). This type of algorithm
 has proved very successful for various SOS models.

\section{Discussion of Results}
\label{SECcomparison}

\subsection*{Ising Model}

\subsubsection*{The Surface Energy}
 We first explain how we obtained estimates for the
 surface energy defined in eq.~(\ref{def_surface_energy}).
 The quantity $\langle H \rangle_{a}$ was always computed
 using either the interface cluster algorithm or the
 local demon algorithm. For the determination of the
 energy with periodic boundary conditions we used
 a Pad\'e approximant of the low
 temperature series for the energy (for details cf.\ \cite{ours}).
 The series was put to
 order 24 in $u=\exp(-4\beta)$ by Bhanot et al.\ \cite{epslow}
 and a little later to order 32 by Vohwinkel \cite{vohwinkel}.
 In \cite{ours}, we found by comparing with Monte Carlo
 results, that the use of the Pad\'e
 approximation was safe for $\beta \geq 0.26$ for $L=8$,
 $\beta \geq 0.24$ for $L=16$ and $L=32$, and for
 $\beta \geq 0.235$ for $L=64$. For the range of couplings
 that we want to focus on in this paper ($\beta \geq 0.35$),
 the use of the Pad\'e approximant is safe (for
 the accuracy required).

 When computing the surface energy per area $\epsilon_s$ one has
 to deal  with
 systematic effects from the finite extension of the
 lattice in $t$- and in $L$-direction.\footnote{For
 a study of finite size effects in a previous high
 precision simulation performed by Ito see \cite{ItoPhysA}}

 First we carefully studied the $t$-dependence of the surface energy
 for three different $\beta$-values, namely
 $\beta=0.45$ (which is in the smooth phase of the interface),
 $\beta=0.4074$ (the roughening point) and
 $\beta=0.3500$ (which is in the rough phase of the interface,
 but still far away from the bulk critical point). The results
 for $\epsilon_s$ for interfaces with extension
 $L=4,8,16,32,64$ and for lattices with various values
 of $t$ are quoted in the tables \ref{tabene1}, \ref{tabene2}
 and \ref{tabene3}. The statistics is as follows: For the
 runs with the cluster algorithm we performed 400 000
 measurements, separated by an update step made up from
 generating and flipping 8 clusters of two different
 types, cf.\ \cite{ours}, and a subsequent Metropolis
 sweep. For the runs with the demon algorithm we
 made 100 000 measurements on each of the 32 independent
 copies, separated by always 5 cycles as described in
 section \ref{SECalgorithms}. For some sets of parameters
 we made runs with two different random number generators
 and found perfect consistency. As a further check we
 compared Monte Carlo results for $L=t=4$ with the
 {\em exact} result for the energy.

 An inspection of the tables reveals that (for fixed $L$) the estimates
 for $\epsilon_s$ are consistent within error bars for
 $t \geq 4$ and $\beta=0.45$.
 For $\beta=0.4074$ we find consistency for $t \geq 5$ if
 $L \leq 32$ and for $t \geq 6$ if $L=64$, cf.\ the discussion
 of finite $t$ effects in section \ref{SECmodels}.
 In this context it might be interesting to note that
 the bulk correlation length $\xi$ which governs the exponential decay
 of the connected correlation function is
 $0.2809$ for $\beta=0.45$. For $\beta=0.4074$, one has
 $\xi=0.3162$, and for $\beta=0.35$ the correlation length
 is $\xi=0.3897$. These estimates are based on
 a Pad\'e evaluation of a low temperature series by
 Arisue and Fujiwara \cite{ArisueFujiwara}.
 Correlation length estimates from the series truncated at $13^{th}$ order
 for $\beta=0.35$ differ from the Pad\'e approximants in the third digit.

 Using the safe values of $t$ found for the smaller $\beta$-values
 we computed the interface energies presented in table
 \ref{tabene4} for the range from $\beta = 0.35$ to
 $\beta = 0.6$ in steps of~$0.01$ or~$0.005$.
 The table also contains estimates for the $L=\infty$
 interface energy from the low temperature series, truncated
 at $17^{th}$ and at $12^{th}$ order, cf.\ the discussion below.

 Concerning the convergence of the surface energies to the
 infinite $L$ limits we make the following observations:
 For $\beta \geq 0.45$ we find that (within the statistical
 accuracy obtained) the estimates converged well, consistent
 with exponentially small $L$-corrections.
 Of course, the convergence becomes better for larger $\beta$.

 In the rough phase the interface energy is expected to
 behave like $ A + B \, L^2 $. We fitted the $\beta=0.35$ data
 for $L\geq 8$ and $t = 13$ with this ansatz and found
 $A=-5.96(47)$ and $B=3.70428(9)$
 with $\chi^2/$dof = 1.135.\footnote{Infinite $L$-extrapolations
 according to the law $A+B\, L^2$ for data in the rough phase
 were also done by Ito \cite{ItoPhysA}}

 At the roughening transition the situation is more difficult.
 We studied the differences $E_s(2L)-E_s(L)$
 for $L \leq 64$ and found that this quantity scales
 down with a factor of at least 3 always when $L$
 is doubled. This gives us confidence that
 $E_s(\infty)-E_s(128)$ is not bigger than
 $E_s(128)-E_s(64) = 0.0022(1)$.


 The estimates for the interface energies can be compared
 with results from the  low temperature expansion of the same quantity.
 The series for the interface tension is
\be\label{sigmaseries}
\sigma = 2\beta +
         \sum_{n=2}^{17} a_n \, u^n + \mbox{$\cal{O}$} (u^{18}) \, .
\ee
 The coefficients up to order $u^9$ in the low temperature
 variable $u=\exp(-4\beta_{\rm Ising})$ were first determined
 by Weeks et al. They can be found in a paper by
 Shaw and Fisher \cite{ShawFisher}. The higher coefficients
 (order $u^{10}$ through $u^{17}$) were computed recently by
 Arisue \cite{ArisueNewSeries}.
 The coefficients $a_n$ are quoted in table \ref{tabcoeff}.
 In order to get $\epsilon_s$ from $\sigma$, one has to
 differentiate eq.~(\ref{sigmaseries}) with respect to $\beta$.
 The result is a power series in $u$ (there is no longer a term
 $\ln(u)$ present). We compared the truncated series and
 its Pad\'e approximants with the Monte Carlo results
 as follows: For order $k$, with $7 \leq k \leq 17$,
 we plotted the result of the
 truncated series (truncated at order $k$) together with the results
 of 4 or 5 close-to-diagonal Pad\'e approximants
 $[m/n]$, with $m+n=k$.
 For even $k$ we took the five Pade\'s with
 $m=k/2-2,\dots,k/2+2$, for odd $k$ we took the four Pad\'es
 with $m=(k-1)/2-1,\dots,(k-1)/2+2$. In all figures
 the truncated series estimates are plotted with a `+' and
 connected with a broken line.
 For the
 Pad\'e estimates we use diamonds.
 On the right hand side of the plots we present our Monte
 Carlo estimates of the surface energy for different lattice
 sizes. The $L=64$ result is also plotted with two vertical
 lines for easier comparison with the series results.

 Figure 1 shows this comparison for $\beta=0.50$
 which is in the smooth phase of the interface and
 far away from the roughening transition.
 For order $\geq 16$ the truncated series and the Pad\'e
 approximants have become consistent with the Monte Carlo
 estimate. Note however, that there is no apparent
 advantage in using the Pad\'e approximants instead
 of the truncated series.

 $\beta=0.45$ is still in the smooth phase.
 The comparison is summarized in figure 2. Notice
 the much larger scale of the y-axis in this plot
 compared to figure 1. The Pad\'e approximants
 scatter a lot, especially around order 13 where
 the series changes its sign. If one looks
 only at order $\leq 12$, the truncated series seems
 to be even superior to the Pad\'e approximations.

 The scenario becomes even more drastic if one
 proceeds to the roughening region.
 In figure 3 we show the comparison of the different
 approximations at the roughening point $\beta=0.4074$.
 Here obviously neither the truncated series nor the
 Pad\'e approximants lead to a reasonable approximation.

\subsubsection*{The Surface Tension}
 The estimates for the surface tension quoted in this paper
 were obtained with the method
 described in \cite{ours}.\footnote{We plan to give
 a comparison of the different methods to determine
 the interface tension (see, e.g.,
 \cite{Klessinger,BergHans,ours,Sanie,direct,ItoPhysA})
 in a forthcoming publication \cite{3dtranseff}}

 However, for the $\beta$-range above $0.35$ we used
 the new and much more precise estimates for the surface
 energy as quoted in table \ref{tabene4}.
 From the results for the surface energy we determined the
 surface tension using the method of `integration over $\beta$'
 as outlined in section \ref{SECmodels}.
 In \cite{ours}
 we used two different starting points $\beta_0$ for the integration,
 namely $\beta_0 \approx \beta_c$ and $\beta_0 = 0.6$. Both
 methods yielded compatible results.
 In table \ref{tabsig}
 we quote our new results obtained by starting the integration
 at the following $\beta$ values:
 For $L=8$ and $L=16$ we took $\beta_0=0.545$, for $L=32$ we
 used $\beta_0 = 0.515$, and for $L=64$ we started the integration
 at $\beta_0 = 0.495$.
 The starting point was determined such that the Monte Carlo
 surface energy estimate and the $17^{th}$ order low temperature
 series for the same quantity are consistent within the present
 error bars.
 Note that the errors quoted in table \ref{tabsig} are statistical
 errors ($1\sigma$ error bars) that do not include systematic effects.
 The estimates rely on fits of the finite $L$ behavior of the
 free energy with the law $C_s + \sigma L^2$.

 For the estimation of systematical errors we used the following
 procedure: One defines
 \be
 \sigma(L) = F_s(L) / L^2 \, .
 \ee
 By the very definition this quantity converges to the
 interface tension in the infinite $L$ limit, however, with
 stronger finite size effects than definition
 eq.~(\ref{roughfs}). So looking at the variation
 of $\sigma(L)$ gives one a feeling of the maximal
 systematic error possible.

 It is also instructive to obtain estimates for $\sigma$ based
 on the law eq.~(\ref{roughfs}), however, using just pairs
 of adjacent $L$-values. Then no fit is needed. In
 table~\ref{tabsigl} we quote these quantities for
 $\beta = 0.402359$ (which corresponds to $u=0.2$) and for
 $\beta = 0.45$. We adopt the following rule for the
 estimation of a systematical error: Take the estimates for
 $\sigma$ from the pair $L=16,32$ and from the pair $L=32,64$
 and compute the difference. Take this as the systematic error
 of the interface tension determined with the fit method.
 For the two examples studied in table~\ref{tabsigl}
 we conclude that for $\beta=0.45$ the systematic
 error is smaller than the statistical error. For
 $\beta = 0.402359$ we arrive at
 $\sigma/(2\beta) = 0.84487(3)$.

 In figures 4 and 5 we show the comparison of the
 Monte Carlo results for the two $\beta$ values
 quoted above\footnote{We chose $u=0.2$ ($\beta = 0.402359$)
 for easier comparison with the work of Shaw and Fisher.
 We compared our Monte Carlo results
 with the series extrapolations also at the more recent
 estimate $\beta_r = 0.4074$, with the same conclusions}
 with the truncated series and
 the Pad\'e approximations.
 The Pad\'es were not performed directly for
 the series for $\sigma$ but (as in \cite{ShawFisher})
 for the quantity $Q(u)=u \exp( 2 \sigma) $.

 The conclusions are similar to the ones for the
 surface energy. Figure 5 demonstrates the
 trap one can get into when the series is too short.
 Shaw and Fisher might have been misled by the convergence
 and consistency of the Pad\'e and differential
 approximants at order 9  and concluded
 that the interface tension could well be approximated
 by Pad\'es of ninth order for temperatures below
 the roughening temperature. The now longer series
 shows that this is a wrong conclusion.
 Note that the Pad\'es seem to converge again at
 the by now highest available order. But still,
 the value is definitely off from our Monte Carlo
 estimate.\footnote{A preliminary Pad\'e analysis for
 the interface tension was already performed
 by Arisue in \cite{ArisueNewSeries}. His results
 are perfectly consistent with ours.
 However, he could not compare with an
 independent Monte Carlo result}

 Shaw and Fisher pointed out that the use of inhomogeneous
 differential approximants \cite{IDA}
 might be superior to using
 Pad\'es. In fact, these approximants generalize
 Pad\'es and are suitable to deal with functions that
 have a critical behavior like $\sim A(x)(x-x_c)^{-\gamma}+B(x)$.
 Note, however, that the singularity of the free energy of the
 Ising interface is not of this type.

 Table \ref{tabida} shows the results obtained by evaluating
 inhomogeneous differential approximants for $\sigma/(2\beta)$
 at $\beta=0.402359$ which corresponds to $u=0.2$.
 Like Shaw and Fisher, we computed the approximant for
 the quantity $Q(u)$ defined above and then took the logarithm.
 We here only discuss the order 9 and order 17 approximants.
 The order 9 approximants are
 fairly stable, however yield too small results (as the
 Pad\'e approximants of this order do).
 Recall that the Monte Carlo estimate for this $\beta$-value
 is $\sigma/(2\beta) = 0.84487(3)$.
 The order 17 approximants are also fairly stable, however, they
 now overshoot the Monte Carlo estimate definitely.
 We conclude that using inhomogeneous differential approximants
 does not cure the problem (as was to be expected).
 The analysis of the surface tension at $\beta=0.4074$ leads
 to a similar result.

 In fig.\ 6 we compare our results for the surface tension with the
 whole range approximant of Fisher et al.\
 \cite{private,ShawFisher,FisherWen} in the $\beta$-range $0.35-0.50$.
 The approximant provided by Fisher and Wen
 \cite{private} is obtained as discussed
 in ref.\ \cite{ShawFisher}. In addition the critical amplitude
 of the surface tension is fixed
 to the value given by Mon \cite{Mon}. The mismatch of the two curves
 can be explained by the failure of the
 approximants to the low temperature
 series  of order 9 at $u=0.20$. The surface tension at $u=0.20$ is
 underestimated and since the surface energy is overestimated the gap
 between the two curves increases with decreasing $\beta$.

 For $\beta$-values close to the bulk
 critical temperature the Monte Carlo result and the interpolation
 are consistent again. This fact confirms the validity of the
 result for the surface tension amplitude obtained by Mon \cite{Mon}.

\subsubsection*{The Specific Heat}

 The specific heat of models undergoing a KT phase transition
 was investigated in a number of Monte Carlo studies.
 For the XY model a peak
 of the specific heat is found, which is located in the
 massive phase of the model at about $0.9 \beta_c$,
 see e.g.\ refs.~\cite{tobo,gupta,ulli}.
 Swendsen \cite{swendsen}
 also found a peak of the specific heat
 for the DGSOS and ASOS models in the massive (smooth)
 phase of the models.
 For the BCSOS model (or F-model) that is an SOS model with
 the constraint that two neighboring height variables
 must differ by +1 or --1, the specific heat can be
 computed exactly (chapter 8 in the book of Baxter \cite{baxter}).
 The peak of the specific heat lies clearly in the massive phase.

 In contrast to these findings, Shaw and Fisher \cite{ShawFisher}
 arrive at the conclusion (based on their series
 analysis) that the peak of the specific heat of the 3D Ising
 interface is located in the massless (rough) phase.

 The position of the specific heat peak is, of course, not a universal
 feature of the KT transition since the specific heat stays finite for
 all temperatures.
 However, the ASOS model is supposed to approximate the Ising
 interface at the roughening transition even quantitatively quite well.
 Hence it would be quite surprising if the specific heat peak of the Ising
 interface should be in the massless phase while that of the
 ASOS model is located in the massive phase.

 We computed the derivative of the surface energy with respect to the
 inverse temperature $\beta$ from finite differences of the energy.
 The results for $L=8,16,32$ and $64$ are given in fig.~7.
 For comparison we give the truncated series result
 for order 12 and 17. For the lattices of size $L=16,32$ and $64$ the
 derivative of the energy clearly exhibits a peak for
 $\beta \sim 0.43 > \beta_r$. The peak of the specific heat
 $C=-\frac{1}{T^2} \frac{dE}{d\beta}$ itself is even slightly deeper in the
 massless phase.
 This is in contradiction to the Shaw and Fisher result. The
 peak height still increases considerable for $L=64$ compared to $L=32$.
 This fact indicates that length scales of order 10 lattice
 spacings are largely involved
 in the generation of the specific heat peak. On the other hand the correlation
 length at $\beta=0.43$ is finite. It should be of order 100.

 For $\beta > 0.46$ the derivative of the energy obtained from the
 $12^{th}$ and $17^{th}$ order truncated series are close together and
reproduce
 the Monte Carlo result within error bars.
 The $12^{th}$ order truncated series contains only coefficients with positive
 sign and is hence increasing monotonically with increasing $u$. Obviously
 it cannot predict the peak of the specific heat.
 The situation is slightly different
 for the  $17^{th}$ order truncated series. The curve displays a peak
 at $\beta=0.3747...$ deep in the massless phase, but obviously wrong.

 One expects the low temperature series of the free energy to converge for
 $\beta > \beta_r$. Hence in principle one should be able to obtain the
 specific heat peak accurately from the truncated series of sufficiently
 high order. However, since length
 scales of more than ten are involved one would have to compute diagrams of
this
 extension.

\subsection*{ASOS Model}

 It is instructive to study the same questions in the SOS approximation.
 The motivation for this is twofold: First, the
 comparison of the results demonstrates that the
 problem studied is indeed due to the Kosterlitz-Thouless nature of
 the roughening transition. Furthermore, e.g.\ in the
 ASOS model it is much easier to get precise data
 for large $L$.
 Table \ref{MCasos} shows our Monte Carlo results
 for the energy of the ASOS model as introduced in
 section \ref{SECmodels}.
 The data were obtained with the help of the
 VMR algorithm. Every single simulation consisted of
 at least 20 000 updates.

 In table \ref{tabcoeff} we quote the coefficients of the low
 temperature expansion
 of the free energy per area for the ASOS model. The expansion
 variable is $u=\exp(-4 \beta_{\rm Ising})$, with
 $\beta_{\rm Ising} = \beta_{\rm ASOS}/2$.
 The series is
\be
 \sigma_{\rm ASOS}
 = 2\beta + \sum_{n=2}^{12} a_n \, u^n + \mbox{$\cal{O}$} (u^{13}) \, .
\ee
 The coefficients
 $a_n$ with $n \leq 9$ are due to Weeks et al. (cited in
 \cite{ShawFisher}). We extended this series to order 12,
 using a technique due to Arisue and Fujiwara
 \cite{ArisueFujiwara,ArisueNewSeries}. Our result for the
 ninth order is at odds with the result cited in \cite{ShawFisher},
 where $a_9=3185/3$ is quoted. We used our own
 value in what follows.

 It is interesting to note that the series (like the corresponding
 Ising series) has a characteristic change of sign in the
 coefficients.
 From the series for the free energy one easily obtains the
 series for the energy per area. Figures 8 and 9 compare
 again our Monte Carlo results with the truncated series and
 with Pad\'e approximants of increasing order.
 The observations are essentially the same as for the
 Ising model. Since we extended the series beyond the
 order where the change of sign takes place we can observe
 the Pad\'e approximants becoming unstable around this
 order.

\section{Conclusions}
\label{SECconclusions}

 In this paper, we presented a comparison of Monte
 Carlo results for interface properties with
 low temperature series. We took the obvious
 discrepancy between the methods as a motivation
 to improve confidence in the Monte Carlo estimates
 by providing a detailed study of possible
 systematic errors.

 The failure of the series approximations to improve the series result
 compared with the truncated series as discussed in
 this article is not completely unexpected.
 However, our detailed study reveals the
 seriousness of the problem and calls
 for new approaches to deal with
 essential singularities in series expansions.

\section*{Acknowledgment}
 We would like to thank M.E. Fisher
 for detailed and stimulating correspondence
 about the subjects discussed in this article.
 We thank H. Arisue for communicating his
 series results prior to publication.
 We thank M.E. Fisher and H. Wen for
 sending us detailed data of their
 interface tension estimates.



\listoftables

\newpage

\vskip3cm
\noindent
{\Large \bf Figure Captions}

\vskip1cm
\noindent
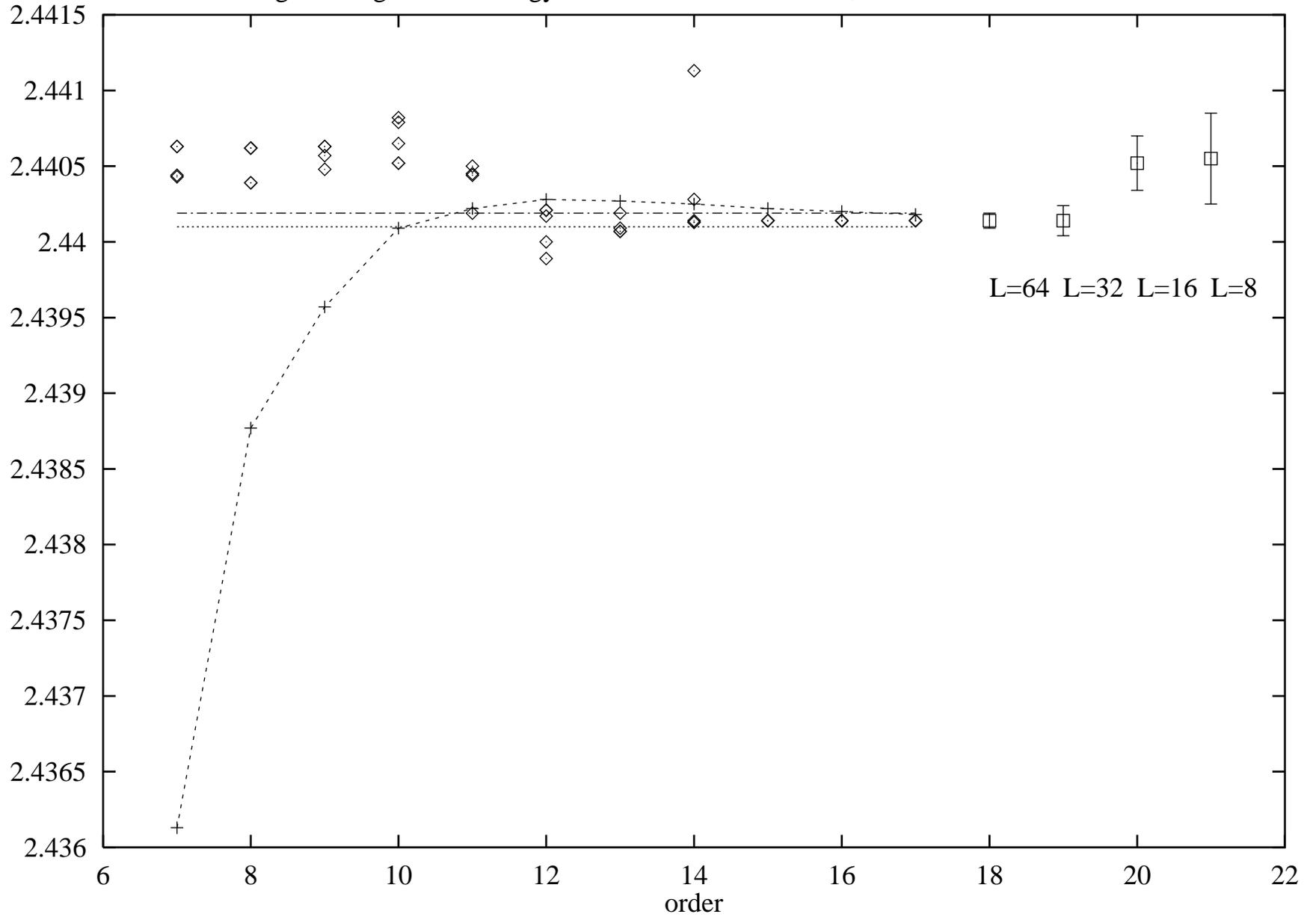
{\bf Fig.\ 1:} Ising surface energy at $\beta=0.5000$ from Pad\'e,
               truncated series and Monte Carlo.
               Truncated series estimates are plotted with a `+' and
               connected with a broken line.
               The close to diagonal Pad\'e estimates are
               given with diamonds.
               The data with error bars present our Monte Carlo
               results for the different lattice sizes.
               The $L=64$ estimate is also given
               by two vertical lines.

\vskip3mm\noindent
{\bf Fig.\ 2:} Ising surface energy at $\beta=0.4500$ from Pad\'e,
               truncated series and Monte Carlo.
               The symbols are the same as in fig.\ 1
\vskip3mm\noindent
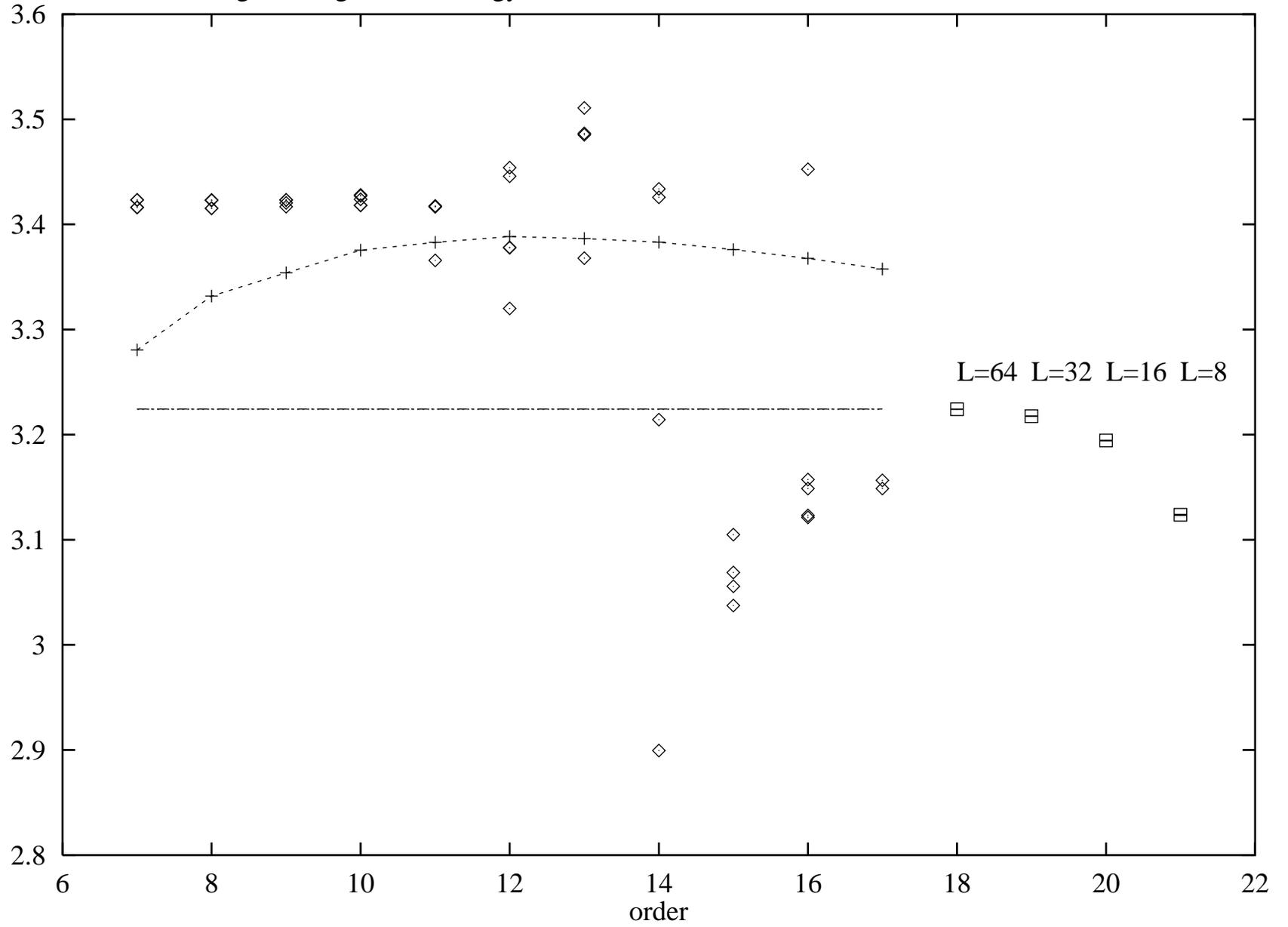
{\bf Fig.\ 3:} Ising surface energy at $\beta=0.4074$ from Pad\'e,
               truncated series and Monte Carlo.
               The symbols are the same as in fig.\ 1

\vskip3mm\noindent
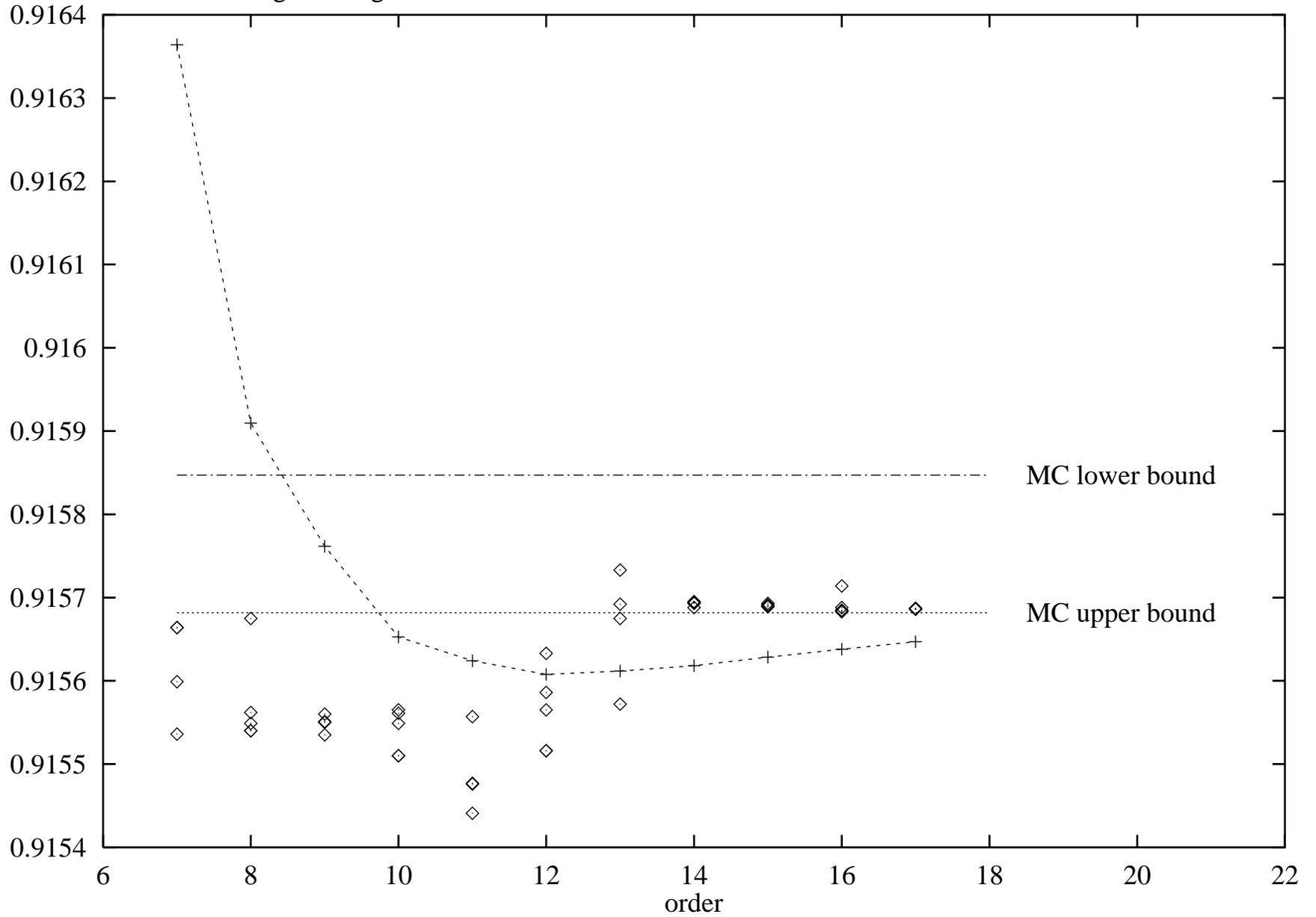
{\bf Fig.\ 4:} Ising surface tension at $\beta=0.4500$ from Pad\'e,
               truncated series and Monte Carlo.
               The symbols are the same as in fig.\ 1

\vskip3mm\noindent
{\bf Fig.\ 5:} Ising surface tension at $u=0.2$ ($\beta=0.4204$) from Pad\'e,
               truncated series and Monte Carlo. For comparison we
               also quote here the Shaw and Fisher estimate
               (based on the ninth order series evaluation).
               The symbols are the same as in fig.\ 1

\vskip3mm\noindent
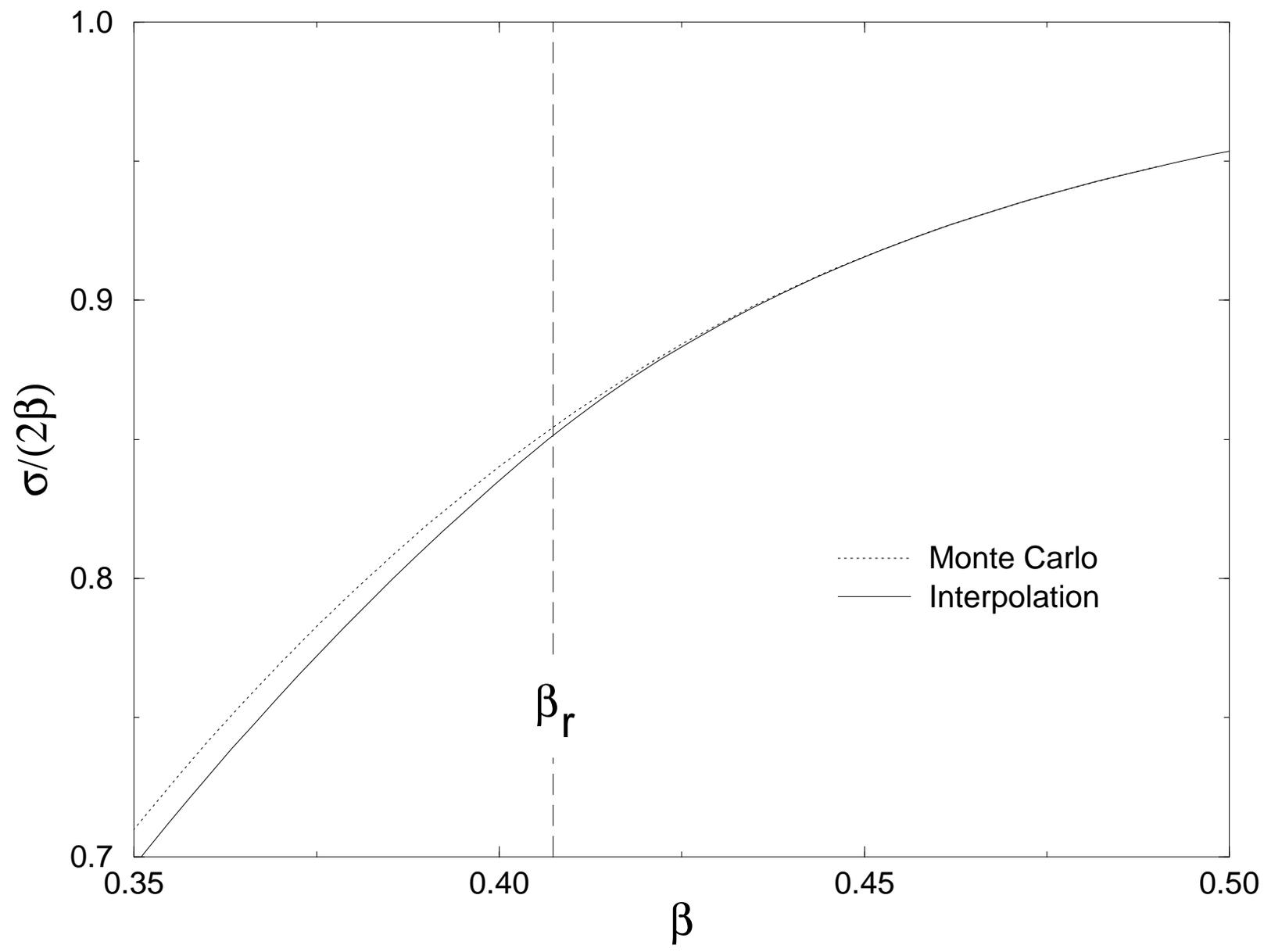
{\bf Fig.\ 6:} The surface tension obtained by Monte Carlo simulation
               (dotted line) is plotted
               as a function of the inverse temperature $\beta$.
               The error bars are smaller than the line width.
               For comparison
               we give the interpolation result of Fisher et al.\
               \cite{private,ShawFisher,FisherWen} (solid line).
               The roughening transition at $\beta_r = 0.4047(3)$ is
               indicated by a vertical dashed line.

\vskip3mm\noindent
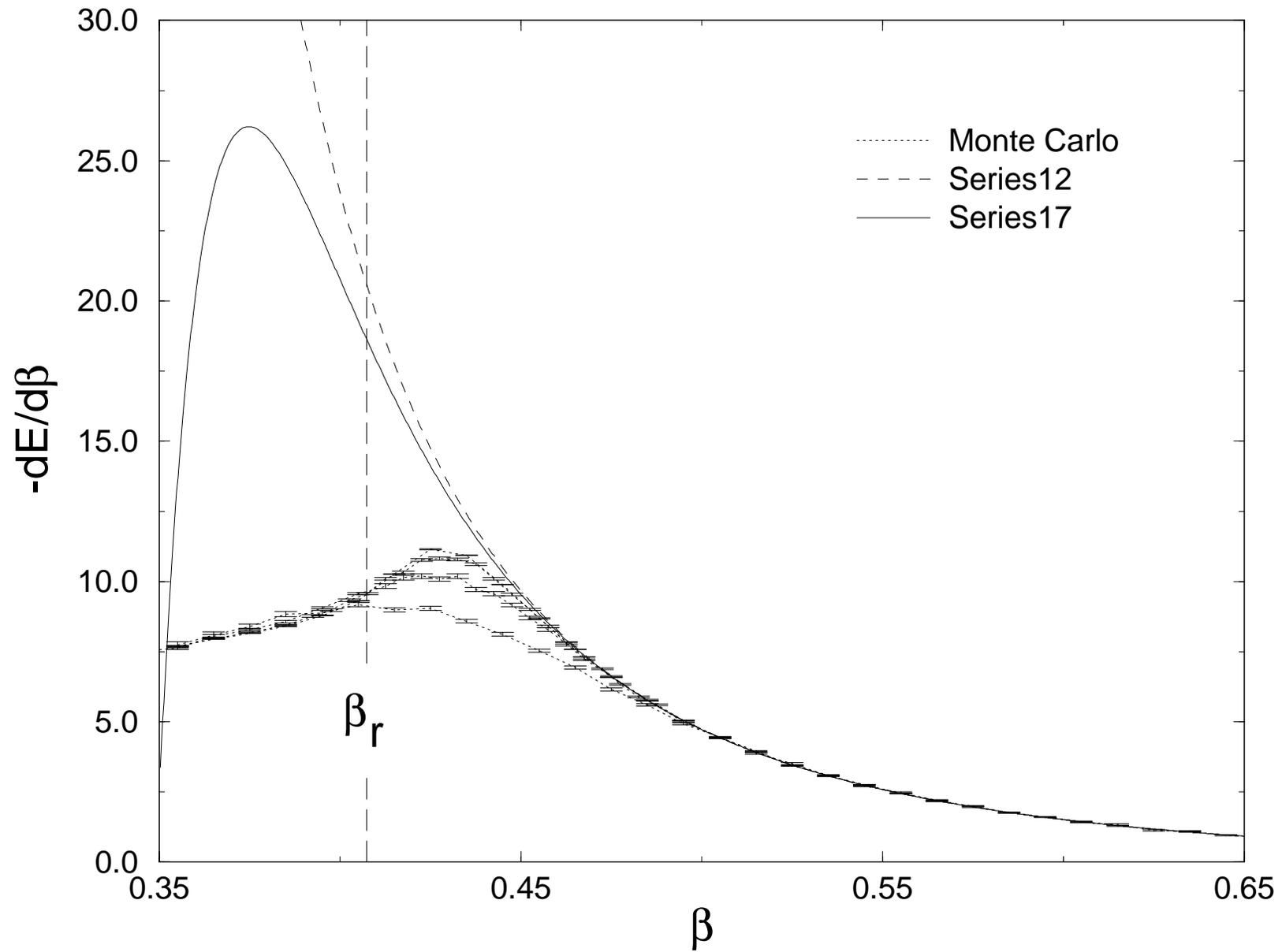
{\bf Fig.\ 7:}
  The derivative of the surface energy with respect to
  $\beta$ is plotted as a function of the inverse temperature $\beta$.
  The dotted lines gives the derivative obtained from finite differences
  of the surface energy for $L=8$,16,32, and 64, computed by
  Monte Carlo.
  The height of the peak grows with increasing $L$.
  For comparison we give the result from the series truncated at
  order 12 (Series12) and truncated at order 17 (Series17).
  The roughening  transition at $\beta_r$ is indicated   by a vertical
  dashed line.

\vskip3mm\noindent
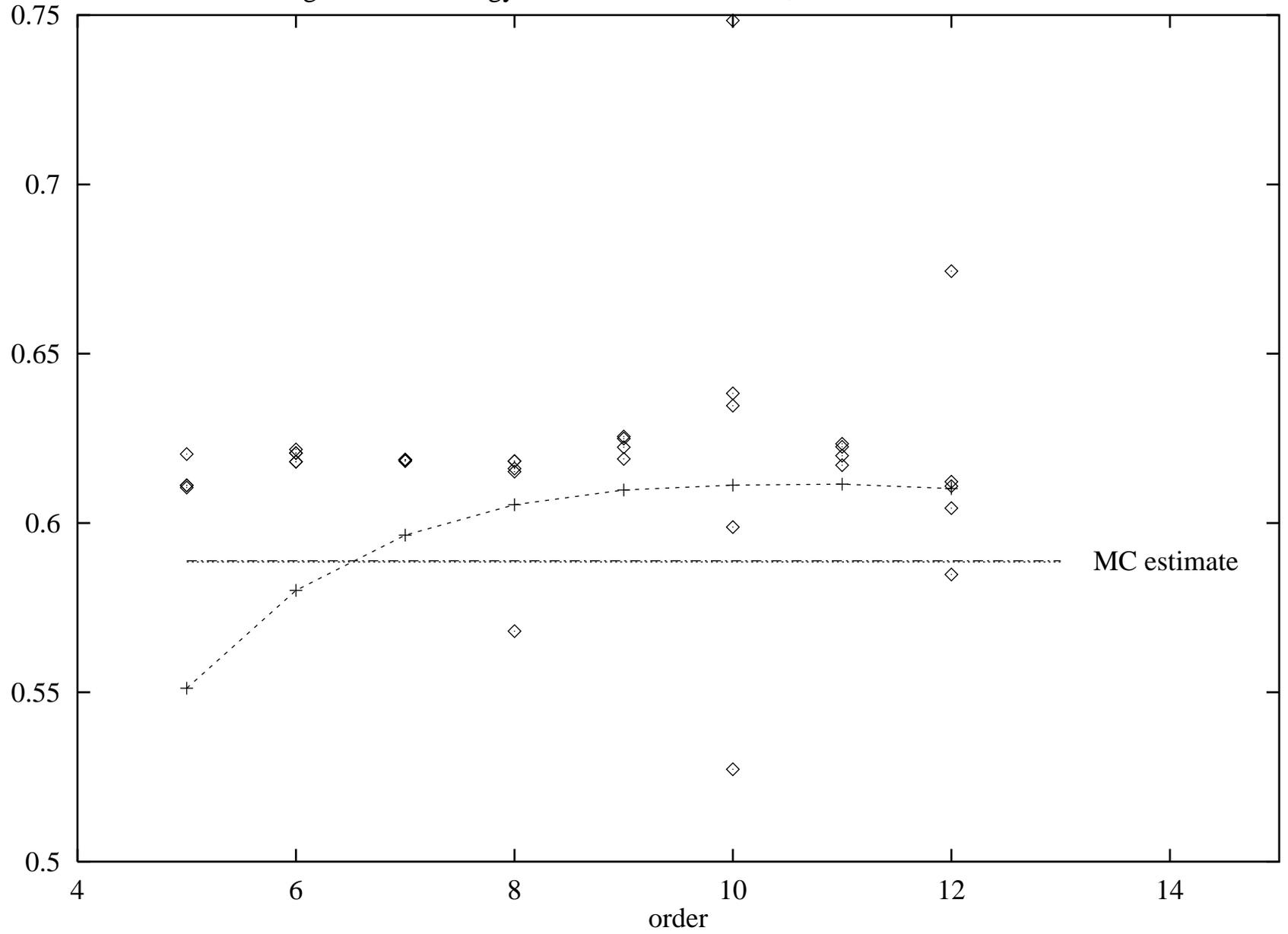
{\bf Fig.\ 8:} ASOS energy at $\beta=0.85$ from Pad\'e,
               truncated series and MC.
               The symbols are the same as in fig.\ 1.

\vskip3mm\noindent
{\bf Fig.\ 9:} ASOS energy at $\beta=0.81$ from Pad\'e,
               truncated series and MC.
               The symbols are the same as in fig.\ 1.


\begin{table}[h]
\small
\begin{center}
\begin{tabular}{|c|c|r|}
\hline
machine & algorithm & \mc{1}{c|}{mspins/sec} \\
\hline
 SUN ELC     & mc & 2.3  \\
 SUN ELC     & c  & 1.1  \\
 SPARC 10    & mc & 7.1  \\
 SPARC 10    & c  & 3.2  \\
 HP 9000/735 & mc & 28.0 \\
 HP 9000/735 & c  & 14.7 \\
\hline
 \end{tabular}
  \parbox[t]{.85\textwidth}
  {
  \caption[Algorithm performance]{\label{tabspeed}
  Performance of the multi-spin program on various workstations,
  given in units of $10^6$ spin updates per second.
  `mc' means micro-canonical run,
  `c' with canonical update of the demons}
  }
\end{center}
\end{table}

\begin{table}
\small
\begin{center}
\begin{tabular}{|c|c|c|r|}
\hline
author(s)           & year & machine & \mc{1}{c|}{mspins/sec}  \\
\hline
 Wansleben et al.  \cite{wansleben}  & 1984
                                     & CYBER 205 (two pipes) &   21.2 \\
 Creutz et al.     \cite{creutz86}   & 1986
                                     & CYBER 205 (two pipes) &  117.0 \\
 Kikuchi and Okabe \cite{kikuchi}    & 1987
                                     & NEC SX-2              &  251.0 \\
 Ito and Kanada    \cite{itokanada}  & 1990
                                     & HITAC S820/80         & 1960.0 \\
 Ito               \cite{itothesis}  & 1991
                                     & VP2600/10             & 2190.0 \\
\hline
 \end{tabular}
  \parbox[t]{.85\textwidth}
  {
  \caption[Performance of algorithms of other authors]
  {\label{tabspeedcomparison}
  Performance of the algorithms of other authors
  on supercomputers for comparison with
  table \ref{tabspeed}}
  }
 \end{center}
 \end{table}

\begin{table}[h]
\small
\begin{center}
\begin{tabular}{|c|l|l|l|l|l|}
\hline
  \mc{1}{|c}{$t$}    &
  \mc{1}{|c}{$L=4$}  &
  \mc{1}{|c}{$L=8$}  &
  \mc{1}{|c}{$L=16$} &
  \mc{1}{|c}{$L=32$} &
  \mc{1}{|c|}{$L=64$} \\
\hline
 2 & 2.67433(63) & 2.78942(42) & 2.83759(30) & 2.85006(21) & 2.85225(12)  \\
 3 & 2.66940(57) & 2.75544(34) & 2.77860(21) & 2.78114(11) & 2.781092(56) \\
 4 & 2.66799(63) & 2.75238(38) & 2.77445(23) & 2.77717(13) & 2.776936(68) \\
 5 & 2.66775(57) & 2.75246(35) & 2.77440(20) & 2.77677(11) & 2.776790(64) \\
 9 & 2.6700(16)  & 2.7517(9)   & 2.7746(5)   &             &              \\
10 & 2.66645(75) & 2.75223(42) & 2.77442(25) & 2.77683(15) & 2.776835(83) \\
17 & 2.6688(19)  & 2.7541(11)  & 2.774(1)    &             &              \\
\hline
 \end{tabular}
  \parbox[t]{.85\textwidth}
  {
  \caption[Ising interface energy at $\beta=0.4500$]{\label{tabene1}
  Ising interface energy at $\beta=0.4500$}
  }
 \end{center}
 \end{table}

\begin{table}
\small
\begin{center}
\begin{tabular}{|c|l|l|l|l|l|}
\hline
  \mc{1}{|c}{$t$}    &
  \mc{1}{|c}{$L=4$}  &
  \mc{1}{|c}{$L=8$}  &
  \mc{1}{|c}{$L=16$} &
  \mc{1}{|c}{$L=32$} &
  \mc{1}{|c|}{$L=64$} \\
\hline
 2 &  2.9542(7) & 3.2408(5)   & 3.42063(27) & 3.48668(12) & 3.492507(59) \\
 3 & 2.9479(7)  & 3.14030(37) & 3.21718(22) & 3.24322(12) & 3.25120(7)   \\
 4 & 2.9419(8)  & 3.1244(4)   & 3.19578(24) & 3.21901(14) & 3.226227(83) \\
 5 & 2.9414(7)  & 3.12342(37) & 3.19428(22) & 3.21745(12) & 3.224334(62) \\
 6 & 2.9427(8)  & 3.1230(5)   & 3.19410(24) & 3.21735(14) & 3.224389(83) \\
 9 & 2.9384(20) & 3.1235(11)  & 3.1936(6)   & 3.21723(30) &              \\
10 &            &             &             &             & 3.224219(88) \\
11 & 2.9520(10) & 3.1237(5)   & 3.19443(28) & 3.21756(16) &              \\
17 & 2.9431(25) & 3.1247(14)  & 3.1941(7)   &             &              \\
33 &            &             & 3.1946(9)   &             &              \\
\hline
 \end{tabular}
  \parbox[t]{.85\textwidth}
  {
  \caption[Ising interface energy at $\beta=0.4074$]{\label{tabene2}
  Ising interface energy at $\beta=0.4074$. We made an additional
  run with $L=128$ and $t=11$. The result for the surface energy
  is  $3.226375(45)$}
  }
 \end{center}
 \end{table}

\begin{table}
\small
\begin{center}
\begin{tabular}{|c|l|l|l|l|l|}
\hline
  \mc{1}{|c}{$t$}    &
  \mc{1}{|c}{$L=4$}  &
  \mc{1}{|c}{$L=8$}  &
  \mc{1}{|c}{$L=16$} &
  \mc{1}{|c}{$L=32$} &
  \mc{1}{|c|}{$L=64$} \\
\hline
 2 & 3.3173(7)  & 3.69719(30) & 3.77344(14) & 3.77574(7)  & 3.775673(34) \\
 3 & 3.3684(8)  & 3.70942(48) & 3.85405(31) & 3.92648(21) & 3.95577(11)  \\
 4 & 3.3479(9)  & 3.62530(50) & 3.70230(29) & 3.72218(16) & 3.726973(76) \\
 5 & 3.3432(9)  & 3.61092(48) & 3.68357(28) & 3.70146(14) & 3.705558(61) \\
 6 & 3.3437(11) & 3.60982(55) & 3.68145(28) & 3.69886(16) & 3.703239(76) \\
 7 & 3.3446(13) & 3.61025(63) & 3.68106(32) & 3.69870(17) & 3.702884(85) \\
 9 & 3.3431(28) & 3.6094(15)  & 3.6816(8)   & 3.69854(39) &              \\
13 & 3.3440(16) & 3.61122(78) & 3.68080(39) & 3.69851(20) & 3.70282(10)  \\
17 & 3.3411(38) & 3.6102(19)  & 3.6836(10)  &             &              \\
\hline
 \end{tabular}
  \parbox[t]{.85\textwidth}
  {
  \caption[Ising interface energy at $\beta=0.3500$]{\label{tabene3}
  Ising interface energy at $\beta=0.3500$}
  }
 \end{center}
 \end{table}

\begin{table}
\small
\begin{center}
\begin{tabular}{|c|l|l|l|l|l|c|}
\hline
  \mc{1}{|c}{$\beta$}   &
  \mc{1}{|c}{$L=8$}     &
  \mc{1}{|c}{$L=16$}    &
  \mc{1}{|c}{$L=32$}    &
  \mc{1}{|c}{$L=64$}    &
  \mc{1}{|c}{$S_{17}$}  &
  \mc{1}{|c|}{$S_{12}$} \\
\hline
    0.350 &  3.61122(78) &
    3.68080(39) &
    3.69851(20) &
    3.70282(11) &
    4.59933 &          5.77307
     \\
    0.360 &  3.53379(70) &
    3.60450(36) &
    3.62157(19) &
    3.62574(9) &
    4.47093 &          5.09037
    \\
    0.370 &  3.45279(67) &
    3.52385(35) &
    3.54189(18) &
    3.54621(9) &
    4.23424 &          4.56179
    \\
    0.380 &  3.36878(64) &
    3.44095(33) &
    3.46002(16) &
    3.46462(8) &
    3.97351 &          4.14707
    \\
    0.390 &  3.28038(59) &
    3.35522(34) &
    3.37550(16) &
    3.38049(7)  &
    3.72531 &          3.81746
    \\
    0.395&    &      &
    3.33163(16) &
                &
      3.61065 &  3.67785
\\
    0.400 &  3.19154(55) &
    3.26458(30) &
    3.28669(16) &
    3.29260(8) &
    3.50320 &          3.55224
    \\
    0.405&    &      &
    3.24002(15)&
               &
     3.40309  &  3.43889
\\
    0.410 &  3.09971(53) &
    3.16889(29) &
    3.19217(16) &
    3.19938(9)  &
    3.31017 &          3.33632
    \\
    0.415 &    &
    3.11972(29) &
    3.14166(16) &
                &
      3.22415  &  3.24326
 \\
    0.420 &  3.00972(50) &
    3.06903(29) &
    3.09000(17) &
    3.09669(11) &
    3.14463 &          3.15860
    \\
    0.425 &    &
    3.01806(28)  &
    3.03614(18)  &
                 &
    3.07117 &    3.08140
 \\
    0.430 &  2.91929(48) &
    2.96759(27) &
    2.98205(18) &
    2.98512(11) &
    3.00334 &          3.01083
    \\
    0.435 &     &
    2.91664(28)  &
    2.92813(18)  &
                 &
     2.94069 &  2.94618
 \\
    0.440 &  2.83342(47) &
    2.86806(27) &
    2.87502(16) &
    2.87575(9)  &
    2.88278 &          2.88681
    \\
    0.445 &     &
    2.82025(26)  &
    2.82450(16)  &
                 &
    2.82923 &  2.83218
 \\
    0.450 &  2.75223(42) &
    2.77442(25) &
    2.77682(15) &
    2.77683(8)  &
    2.77965 &          2.78182
    \\
    0.455 &     &
   2.73093(23)   &
   2.73180(14)   &
                 &
   2.73370 &  2.73529
 \\
    0.460 &  2.67688(41) &
    2.68951(23) &
    2.68978(14) &
    2.68991(7)  &
    2.69107 &          2.69224
    \\
    0.465 &     &
   2.65063(21)   &
   2.65065(13)   &
                 &
    2.65148 &  2.65234
\\
    0.470 &  2.60753(38) &
    2.61440(20) &
    2.61426(13) &
    2.61408(6)  &
    2.61465 &          2.61528
    \\
    0.475 &     &    &
   2.57979(12)   &
                 &
    2.58036 &  2.58083
 \\
    0.480 &  2.54599(36) &
    2.54823(19) &
    2.54813(12) &
    2.54813(6)  &
    2.54840 &          2.54874
    \\
    0.485 &     &    &
    2.51866(11)  &
                 &
    2.51856 &  2.51882
 \\
    0.490 &  2.48992(32) &
    2.49041(18) &
    2.49060(10) &
    2.49054(5)  &
    2.49068 &          2.49087
    \\
    0.500 &  2.44055(30) &
    2.44052(18) &
    2.44014(10) &
    2.44014(5)  &
    2.44018 &          2.44028
    \\
    0.510 &  2.39611(28) &
    2.39613(18) &
    2.39578(9) &
    &
    2.39580 &          2.39586
    \\
    0.520 &  2.35725(25) &
    2.35658(17) &
    2.35657(8) &
    &
    2.35665 &          2.35668
    \\
    0.530 &  2.32212(23) &
    2.32219(16) &
    2.32208(8)  &
    &
    2.32198 &          2.32200
    \\
    0.540 &  2.29135(23) &
    2.29123(14) &
    2.29129(7)  &
    &
    2.29118 &          2.29119
    \\
    0.550 &  2.26399(20) &
    2.26403(14) &
    2.26363(7)  &
    &
    2.26374 &          2.26375
    \\
    0.560 &  2.23930(20) &
    2.23943(13) &
    2.23904(7)  &
    &
    2.23922 &          2.23923
    \\
    0.570 &  2.21751(19) &
    2.21728(12) &
    2.21724(6) &
    &
    2.21726 &          2.21726
    \\
    0.580 &  2.19748(19) &
    2.19757(11) &
    &
    &
    2.19755 &          2.19755
     \\
    0.590 &  2.17986(17) &
    2.17989(11) &
    &
    &
    2.17981 &          2.17981
    \\
    0.600 &  2.16378(17) &
    2.16387(10) &
    &
    &
    2.16383 &          2.16383
    \\
\hline
 \end{tabular}
  \parbox[t]{.85\textwidth}
  {
  \caption[Ising interface energy for $\beta=0.35-0.60$]
  {\label{tabene4}
  Ising interface energy, Monte Carlo and truncated low temperature series
  results ($S_{i}$ corresponds to truncation at $i^{th}$ order)}
  }
\end{center}
\end{table}

\begin{table}
\small
\begin{center}
\begin{tabular}{|c c r||c c r||c c r|}
\hline
  $\beta$  & $\sigma/(2 \beta)$ & \mc{1}{c||}{X} &
  $\beta$  & $\sigma/(2 \beta)$ & \mc{1}{c||}{X} &
  $\beta$  & $\sigma/(2 \beta)$ & \mc{1}{c|}{X} \\
\hline
.225 & .01629(30)  & 2.3 & .315 & .575804(38)  &  7.0 & .405 & .8499891(27) &
6.2 \\
     & .01622(25)  & 2.0 &       & .575824(38)  &  5.8 &       & .8499979(26) &
21.5 \\
\hline
.230 & .04929(25)  & 4.8 & .320 & .597570(31)  &  6.9 & .410 & .8592883(27) &
2.1 \\
     & .04921(20)  & 3.0 &       & .597585(35)  &  6.5 &       & .8592906(22) &
 3.2 \\
\hline
.235 & .08565(18)  & 1.9 & .325 & .618399(28)  &  5.6 & .415 & .8680694(26) &
2.0 \\
     & .08562(20)  & 1.5 &       & .618420(25)  &  8.1 &       & .8680646(22) &
 7.8 \\
\hline
.240 & .12262(18)  & 1.1 & .330 & .638353(26)  &  7.4 & .420 & .8763354(26) &
10.0 \\
     & .12263(16)  & 1.1 &       & .638390(25)  & 11.3 &       & .8763233(21) &
49.7 \\
\hline
.245 & .15971(15)  & 1.4 & .335 & .657448(21)  &  4.9 & .425 & .8840889(24) &
19.2 \\
     & .15972(14)  & 1.2 &       & .657500(16)  & 15.1 &       & .8840714(19)
&117.0 \\
\hline
.250 & .19623(14)  & 1.2 & .340 & .675699(13)  &  2.8 & .430 & .8913357(25) &
24.3 \\
     & .19627(12)  & 1.2 &       & .675751(12)  & 23.6 &       & .8913154(18)
&178.7 \\
\hline
.255 & .23200(12)  & 1.4 & .345 & .6931576(53) &  3.7 & .435 & .8980930(21) &
13.9 \\
     & .23202(12)  & 1.5 &       & .6931802(49) & 39.6 &       & .8980731(16)
&195.0 \\
\hline
.260 & .266799(93) & 1.7 & .350 & .7098453(66) &  1.0 & .440 & .9043836(18) &
5.2 \\
     & .266831(94) & 1.8 &       & .7098713(62) & 34.5 &       & .9043665(16)
&175.3 \\
\hline
.265 & .300562(97) & 1.5 & .355 & .7258015(44) &  2.5 & .445 & .9102364(18) &
2.0 \\
     & .300592(92) & 2.2 &       & .7258219(36) & 43.5 &       & .9102225(14)
&137.8 \\
\hline
.270 & .333177(90) & 2.4 & .360 & .7410463(39) &  5.6 & .450 & .9156833(14) &
.9 \\
     & .333196(80) & 2.6 &       & .7410681(34) & 51.9 &       & .9156726(12) &
97.3 \\
\hline
.275 & .364607(73) & 2.9 & .365 & .7556038(42) &  5.1 & .455 & .9207540(12) &
1.6 \\
     & .364639(82) & 2.7 &       & .7556250(33) & 59.9 &       & .9207468(12) &
61.3 \\
\hline
.280 & .394894(71) & 1.7 & .370 & .7694993(37) &  4.7 & .460 & .9254779(11) &
1.6 \\
     & .394923(66) & 2.3 &       & .7695206(34) & 64.3 &       & .9254730(10) &
36.2 \\
\hline
.285 & .424033(66) & 1.8 & .375 & .7827565(36) &  6.7 & .465 & .9298820(12) &
1.3 \\
     & .424051(60) & 2.3 &       & .7827780(30) & 68.5 &       & .9298786(10) &
20.3 \\
\hline
.290 & .451998(61) & 1.8 & .380 & .7953967(37) &  8.4 & .470 & .9339903(10) &
2.1 \\
     & .452023(60) & 2.6 &       & .7954174(32) & 70.4 &       & .9339883(09) &
 9.2 \\
\hline
.295 & .478878(63) & 2.2 & .385 & .8074397(33) &  9.2 & .475 & .9378262(08) &
1.3 \\
     & .478887(55) & 2.6 &       & .8074602(31) & 69.4 &       & .9378251(07) &
 3.9 \\
\hline
.300 & .504651(53) & 2.8 & .390 & .8189052(34) & 10.6 & .480 & .9414105(06) &
1.0 \\
     & .504664(52) & 2.9 &       & .8189246(26) & 66.9 &       & .9414100(06) &
 2.5 \\
\hline
.305 & .529376(47) & 3.8 & .395 & .8298090(32) & 10.5 & .485 & .9447624(04) &
1.6 \\
     & .529393(46) & 3.5 &       & .8298263(28) & 59.4 &       & .9447621(03) &
 1.8 \\
\hline
.310 & .553076(44) & 5.2 & .400 & .8401660(29) & 10.1 & .490 & .9478999(04) &
1.8 \\
     & .553089(42) & 4.3 &       & .8401801(25) & 43.6 &       & .9478997(03) &
 1.1 \\
\hline
 \end{tabular}
  \parbox[t]{.85\textwidth}
  {
  \caption[Monte Carlo results for the Ising interface tension]
  {\label{tabsig}
  Monte Carlo results for the interface tension. We always quote
  two numbers. For the upper number the $L=16,32,64$ data were included.
  For lower number also the $L=8$ data were included.
  X denotes $\chi^2$ per degree of freedom for the fit of the
  free energies with $C_s + \sigma L^2$}
  }
\end{center}
\end{table}

\begin{table}[h]
\small
\begin{center}
\begin{tabular}{|r l|r l|}
\hline
\mc{4}{|c|}{$\beta = 0.402359$ ($u=0.2$)}       \\
\hline
  $L$ & $\sigma(L)/(2\beta)$ & pair & $\sigma/(2\beta)$  \\
\hline
  8 & 0.849503(17)         &   8-16 & 0.844959  \\
 16 & 0.8460946(87)        &  16-32 & 0.844882  \\
 32 & 0.8451848(42)        &  32-64 & 0.844858  \\
 64 & 0.8449398(30)        &        &           \\
\hline
\hline
\mc{4}{|c|}{$\beta = 0.45$}                     \\
\hline
  $L$ & $\sigma(L)/(2\beta)$ & pair& $\sigma/(2\beta)$ \\
\hline
  8 & 0.916039(10)         &   8-16 & 0.915567 \\
 16 & 0.9156853(56)        &  16-32 & 0.915683 \\
 32 & 0.9156838(24)        &  32-64 & 0.915683 \\
 64 & 0.9156833(14)        &        &           \\
\hline
  \end{tabular}
   \parbox[t]{.85\textwidth}
   {
   \caption[The quantities $\sigma(L)$]{\label{tabsigl}
   Surface tension $\sigma(L)/(2\beta)$
   together with estimates for the
   same quantity obtained from surface free energies
   for pairs of $L$-values}
   }
 \end{center}
 \end{table}

\begin{table}[h]
\begin{center}
\begin{tabular}{|r|l|l|}
\hline
\mc{1}{|c|}{$n$} & \mc{1}{c|}{$a_n^{\rm Ising}$}
& \mc{1}{c|}{$a_n^{\rm ASOS}$}\\
\hline
 2  & -- 2              & -- 2         \\
 3  & -- 2              & -- 4         \\
 4  & -- 10             & -- 10        \\
 5  & -- 16             & -- 24        \\
 6  & -- 242/3          & -- 194/3     \\
 7  & -- 150            & -- 172       \\
 8  & -- 734            & -- 452       \\
 9  & -- 4334/3         & -- 3184/3    \\
10  & -- 32122/5        & -- 8862/5    \\
11  & -- 10224          & + 1712       \\
12  & -- 106348/3       & + 116804/3   \\
13  & + 53076           &              \\
14  & + 3491304/7       &              \\
15  & + 74013814/15     &              \\
16  & + 27330236        &              \\
17  & + 160071418       &              \\
\hline
 \end{tabular}
  \parbox[t]{.85\textwidth}
  {
  \caption[Low temperature series coefficients $a_n$]
  {\label{tabcoeff}
  Coefficients of the low temperature series for the 3D Ising
  and ASOS interface tension}
  }
\end{center}
\end{table}

\begin{table}[h]
\small
\begin{center}
\begin{tabular}{|cc|cc|cc|cc|}
\hline
\mc{2}{|c}{order 17} &
\mc{2}{|c}{order 15} &
\mc{2}{|c}{order 12} &
\mc{2}{|c|}{order 9} \\
\hline
$[9/5,1]$ & --       & $[7/5,1]$ & 0.849 19 & $[6/3,1]$ & 0.841 35 & $[4/2,1]$
& 0.840 73 \\
$[8/6,1]$ & 0.845 08 & $[6/6,1]$ & 0.846 21 & $[5/4,1]$ & 0.843 64 & $[3/3,1]$
& 0.840 69 \\
$[7/7,1]$ & --       & $[5/7,1]$ & 0.846 18 & $[4/5,1]$ & 0.841 26 & $[2/4,1]$
& --        \\
$[6/8,1]$ & --        &           &          & $[3/6,1]$ & 0.841 25 &
&          \\
\hline
$[9/4,2]$ & 0.846 06 & $[7/4,2]$ & --        & $[5/3,2]$ & 0.841 45 & $[3/2,2]$
& 0.840 41 \\
$[8/5,2]$ & 0.845 66 & $[6/5,2]$ & 0.844 80 & $[4/4,2]$ & 0.841 43 & $[2/3,2]$
& 0.840 29 \\
$[7/6,2]$ & 0.845 75 & $[5/6,2]$ & 0.846 82 & $[3/5,2]$ & 0.841 15 &
&          \\
$[6/7,2]$ & 0.845 45 & $[4/7,2]$ & 0.844 75 &           &          &
&          \\
$[5/8,2]$ & 0.845 92 &           &          &           &          &
&          \\
$[4/9,2]$ & 0.845 90 &           &          &           &          &
&          \\
\hline
$[7/5,3]$ & --        & $[6/4,3]$ & --        & $[5/2,3]$ & 0.841 67 &
$[3/1,3]$
& 0.840 70 \\
$[6/6,3]$ & --        & $[5/5,3]$ & 0.845 79 & $[4/3,3]$ & --        &
$[2/2,3]$
& 0.840 25 \\
$[5/7,3]$ & 0.845 49 & $[4/6,3]$ & 0.845 52 & $[3/4,3]$ & 0.841 74 & $[1/3,3]$
& 0.840 32 \\
$[4/8,3]$ & --        &           &          & $[2/5,3]$ & 0.842 06 &
&          \\
\hline
$[7/4,4]$ & --        & $[6/3,4]$ & --        & $[4/2,4]$ & --        &
&          \\
$[6/5,4]$ & 0.845 55 & $[5/4,4]$ & --        & $[3/3,4]$ & 0.842 96 &
&          \\
$[5/6,4]$ & 0.845 49 & $[4/5,4]$ & --        & $[2/4,4]$ & 0.842 08 &
&          \\
$[4/7,4]$ & --        & $[3/6,4]$ & 0.844 90 &           &          &
&          \\
\hline
$[7/3,5]$ & 0.845 63 & $[5/3,5]$ & --        &           &          &
&          \\
$[6/4,5]$ & 0.845 68 & $[4/4,5]$ & --        &           &          &
&          \\
$[5/5,5]$ & 0.845 63 & $[3/5,5]$ & 0.845 54 &           &          &
&          \\
$[4/6,5]$ & --        &           &          &           &          &
&          \\
$[3/7,5]$ & 0.845 66 &           &          &           &          &
&          \\
\hline
  \end{tabular}
   \parbox[t]{.85\textwidth}
   {
   \caption[Inhomogeneous Differential Approximants]{\label{tabida}
   Surface tension $\sigma/(2\beta)$ at $u=0.2$ ($\beta=0.4024$)
   estimated from inhomogeneous differential
   approximants $[N/M,L]$ for $Q(u)$. The symbol `--' means that the
   corresponding approximant could not be computed because
   it did not exist or the numerics was unstable}
   }
\end{center}
\end{table}

\begin{table}
\begin{center}
\begin{tabular}{|c|l|l|l|l|l|}
\hline
$\beta_{ASOS}$       &
\mc{1}{|c}{$L=64$}   &
\mc{1}{|c}{$L=128$}  &
\mc{1}{|c|}{$L=256$} \\
 \hline
0.81  & 0.70741(57) & 0.70947(33) & 0.70964(18) \\
0.82  & 0.67948(63) & 0.68052(36) & 0.68106(20) \\
0.83  & 0.64976(68) & 0.65157(40) & 0.65182(20) \\
0.84  & 0.64976(68) & 0.62025(37) & 0.62048(20) \\
0.85  & 0.5874(8)   & 0.58837(39) & 0.58869(19) \\
0.87  & 0.5272(6)   & 0.52615(35) & 0.52643(14) \\
0.90  & 0.4420(6)   & 0.44255(26) & 0.44245(14) \\
0.95  & 0.3320(6)   & 0.33177(28) & 0.33173(15) \\
1.00  & 0.2517(4)   & 0.25153(23) & 0.25128(10) \\
1.10  & 0.1496(3)   & 0.14869(15) & 0.14905(7)  \\
\hline
\end{tabular}
  \parbox[t]{.85\textwidth}
  {
  \caption[Monte Carlo results for the energy of the ASOS model]
  {\label{MCasos}
  Monte Carlo results for the energy of the ASOS model}
  }
\end{center}
\end{table}


\begin{thebibliography}{99}

\bibitem{martinthesis}
 M. Hasenbusch,
 PhD thesis, Universit\"at Kaiserslautern, 1992.

\bibitem{betac}
 C.F. Baillie, R. Gupta, K.A. Hawick and G.S. Pawley, \\
 Phys.\ Rev.\ B 45 (1992) 10438.

\bibitem{KT}
 J.M. Kosterlitz and D.J. Thouless,
 J. Phys.\ C 6 (1973) 1181; \\
 J.M. Kosterlitz,
 J. Phys.\ C 7 (1974) 1046.

\bibitem{ShawFisher}
 L.J.\ Shaw and M.E.\ Fisher,
 Phys.\ Rev.\ A 39 (1989) 2189.

\bibitem{ArisueNewSeries}
 H. Arisue, preprint OPCT 93-1, May 1993.

\bibitem{Binder}
 K. Binder (ed.),
 {\em Monte Carlo Methods in Statistical Physics}, \\
 Springer, Berlin 1986.

\bibitem{Klessinger}
 S. Klessinger, G. M\"unster,
 Nucl.\ Phys.\ B 386 (1992) 701.

\bibitem{BergHans}
 B.A. Berg, U. Hansmann and  T. Neuhaus,
 Z. Phys.\ B 90 (1993) 229.

\bibitem{ours}
 M. Hasenbusch and K. Pinn,
 Physica A 192 (1993) 342.

\bibitem{Sanie}
 H. Gausterer, J. Potvin, C. Rebbi and S. Sanielevici, \\
 Physica A 192 (1993) 525.

\bibitem{direct}
 M. Hasenbusch,
 J. Phys.\ I France 3 (1993) 753.

\bibitem{ItoPhysA}
 N. Ito, Physica A 196 (1993) 591.

\bibitem{VMRthegang}
 H.G. Evertz, M. Marcu, K. Pinn and S. Solomon,
 Phys.\ Lett.\ B 254 (1991) 185.

\bibitem{PrivmanFisher}
 V. Privman and M.E. Fisher,
 J. Stat.\ Phys. 33 (1983) 385.

\bibitem{KTSOS}
 S.T. Chui and J.D. Weeks,
 Phys.\ Rev.\ B 14 (1976) 4978; \\[1mm]
 T. Ohta and K. Kawasaki,
 Prog.\ Theor.\ Phys.\ 60 (1978) 365; \\[1mm]
 For reviews on SOS models see, e.g.: \\[1mm]
 D.B. Abraham,
 {\em Surface Structures and Phase Transitions -- Exact Results\/},
 in {\em Phase Transitions and Critical Phenomena\/} Vol.\ 10,
 C. Domb and J.L. Lebowitz (eds.), Academic, 1986, p.\ 1;\\[1mm]
 H. van Beijeren and I. Nolden, {\em The Roughening Transition\/}, in
 {\em Topics in Current Physics, Vol.\ 43: Structure and
 Dynamics of Surfaces~II\/}, W. Schommers and
 P. von Blanckenhagen (eds.), Springer, 1987, p.\ 259;\\[1mm]
 J. Adler, Physica A 168 (1990) 646.

\bibitem{Borgsetal}
 C. Borgs and J.Z. Imbrie,
 Comm.\ Math.\ Phys.\ 145 (1992) 235

\bibitem{PrivmanCapillary}
 V. Privman,
 Int.\ J.\ Mod.\ Phys.\ C 3 (1992) 857.

\bibitem{brezin}
 E. Br\'ezin and J. Zinn-Justin,
 Nucl.\ Phys.\ B 257 (1985) 867.

\bibitem{munster}
 G. M\"unster,
 Nucl.\ Phys.\ B 340 (1990) 559.

\bibitem{bunk}
 B. Bunk,
 Int.\ J. Mod.\ Phys.\ C 3 (1992) 889.

\bibitem{cas}
 M. Caselle, F. Gliozzi and S. Vinti,
 Phys.\ Lett.\ B 302 (1993) 74.

\bibitem{gelfand}
  M.P. Gelfand and M.E. Fisher,
  Physica A 166  (1990) 1.

\bibitem{matching}
 M. Hasenbusch, M. Marcu and K. Pinn, \\
 Nucl.\ Phys.\ B (Proc.\ Suppl.) 26 (1992) 598, and in preparation.

\bibitem{InterfaceAlgorithm}
 M. Hasenbusch and S. Meyer,
 Phys.\ Rev.\ Lett.\ 66 (1991) 530.

\bibitem{creutz83}
 M. Creutz,
 Phys.\ Rev.\ Lett.\ 50 (1983) 1411.

\bibitem{creutz84}
 M. Creutz, G. Bhanot and H. Neuberger,
 Nucl.\ Phys.\ B235 [FS11] (1984) 417.

\bibitem{creutz86}
 M. Creutz, K.J.M. Moriarty and M. O'Brien, \\
 Comput.\ Phys.\ Comm.\ 42 (1986) 191.

\bibitem{kari93}
 K.\ Rummukainen,
 Nucl.\ Phys.\ B 390 (1993) 621.

\bibitem{epslow}
 G.\ Bhanot, M.\ Creutz and J.\ Lacki,
 Phys.\ Rev.\ Lett.\ 69 (1992) 1841.

\bibitem{vohwinkel}
 C.\ Vohwinkel,
 Phys.\ Lett.\ B 301 (1993) 208, and private communication.

\bibitem{ArisueFujiwara}
 H. Arisue and T. Fujiwara,
 Nucl.\ Phys.\ B 285 [FS 19] (1987) 253; \\
 H. Arisue and T. Fujiwara,
 preprint RIFP-588, 1985, unpublished.

\bibitem{3dtranseff}
 M. Hasenbusch, K. Pinn and K. Rummukainen, in preparation

\bibitem{IDA}
 M.E. Fisher and H. Au-Yang,
 J. Phys.\ A 12 (1979) 1677.

\bibitem{private}
 M.E. Fisher, private communication.

\bibitem{FisherWen}
 M.E. Fisher and H. Wen,
 Phys.\ Rev.\ Lett.\ 68 (1992) 3654.

\bibitem{Mon}
 K.K. Mon,
 Phys.\ Rev.\ Lett.\ 60 (1988) 2749.

\bibitem{tobo}
 J. Tobochnik and G.V. Chester,
 Phys.\ Rev.\ B 20 (1979) 3761.

\bibitem{gupta}
 R. Gupta, J. Delapp, G.G. Batrouni and G. Fox, \\
 Phys.\ Rev.\ Lett.\ 61 (1988) 1996.

\bibitem{ulli}
 U. Wolff,
 Nucl.\ Phys.\ B 322 (1989) 759.

\bibitem{swendsen}
 R.H. Swendsen,
 Phys.\ Rev.\ B 15 (1977) 5421.

\bibitem{baxter}
 R. Baxter,
 {\em Exactly Solved Models in Statistical Mechanics},
 Academic 1982.

\bibitem{wansleben}
 S. Wansleben, J.G. Zabolitzky and C. Kalle,
 J. Stat.\ Phys.\ 37 (1984) 283.

\bibitem{kikuchi}
 M. Kikuchi and Y. Okabe,
 Phys.\ Rev.\ B 35 (1987) 5382.

\bibitem{itokanada}
 N. Ito and Y. Kanada,
 {\em Proceedings of Supercomputing '90},
 IEEE Computer Society Press, 1990, Los Alamitos, p.\ 753.

\bibitem{itothesis}
 N. Ito, private communication.

\end{thebibliography}
\end{document}